\documentclass{emulateapj} 
\usepackage{apjfonts}

\begin{document}

\title{Non-Thermal emission from the photospheres of Gamma-Ray Burst
  outflows.\\ I: High frequency tails.}

\shorttitle{Non-thermal GRB photospheres I}

\author{Davide Lazzati\altaffilmark{1}, Mitchell
  C. Begelman\altaffilmark{2,3}} \shortauthors{Lazzati \& Begelman}

\altaffiltext{1}{Department of Physics, NC State University, 2401
Stinson Drive, Raleigh, NC 27695-8202}


\altaffiltext{2}{JILA, University of Colorado, 440 UCB, Boulder, CO
80309-0440}

\altaffiltext{3}{University of Colorado, Department of Astrophysical and
Planetary Sciences, 389 UCB, Boulder, CO 80309-0389}

\begin{abstract} 
  We study the spectrum of high frequency radiation emerging from
  mildly dissipative photospheres of long-duration gamma-ray burst
  outflows. Building on the results of recent numerical
  investigations, we assume that electrons are heated impulsively to
  mildly relativistic energies by either shocks or magnetic
  dissipation at Thomson optical depths of several and subsequently
  cool by inverse Compton, scattering off the thermal photons of the
  photosphere. We show that even in the absence of magnetic field and
  non-thermal leptons, inverse Compton scattering produces power-law
  tails that extend from the peak of the thermal radiation, at several
  hundred keV, to several tens of MeV, and possibly up to GeV
  energies. The slope of the high-frequency power-law is predicted to
  vary substantially during a single burst, and the model can easily
  account for the diversity of high-frequency spectra observed by
  BATSE. Our model works in baryonic as well as in magnetically
  dominated outflows, as long as the magnetic field component is not
  overwhelmingly dominant.
\end{abstract}

\keywords{gamma-ray: bursts --- radiation mechanisms: non-thermal ---
  methods: numerical --- relativity}

\section{Introduction}

Understanding the origin of the prompt emission of Gamma-Ray Bursts
(GRBs) has been hampered by two major challenges that have proven
formidable obstacles despite more than four decades of
studies. Radiating photons from an ultra-relativistic jet requires
first a mechanism to convert at least some of the bulk kinetic energy
into thermal motion of electrons, second a mechanism to radiate the
electrons' energy into photons. These two processes are known as the
problem of the dissipation mechanism and of the radiation process.

The standard model that has been developed over the years assumes that
internal shocks are responsible for the dissipation (Rees \& M{\'e}sz{\'a}ros
1994), while electrons gyrating in a shock-generated magnetic field
produce the radiation via the synchrotron process (M{\'e}sz{\'a}ros et
al. 1994; Piran 1999; Lloyd \& Petrosian 2000). Both mechanisms are
fraught with numerous problems.

Internal shocks assume that the outflow is released by the central
engine with fluctuations in the Lorentz factor, so that different
parts of the flow collide with each other, producing shocks that
dissipate energy. Unfortunately, the internal shock mechanism has very
little predictive power, since any behavior of the light curve can be
explained by a suitable choice of the ejection history of the central
engine. Since the physics of the emission from the engine is largely
unconstrained, it is very hard to disprove internal shocks based on
any observation. The only robust prediction of internal shocks as a
dissipation mechanism is the low efficiency of the process (Kobayashi
et al. 1997; Lazzati et al. 1999; Spada et al. 2000), due to the fact
that internal shocks can dissipate only the energy associated with
differential motions and not the energy associated to the bulk motion,
which is much larger. Observations, instead, show that at least for a
fraction of bursts, the efficiency of the prompt phase is 50 per cent,
if not higher (Zhang et al.  2007). An alternative model that is
becoming increasingly popular is magnetic dissipation in a Poynting
dominated outflow (e.g., Thompson 1994; Spruit et al. 2001; Giannios
\& Spruit 2005). Even though magnetic dissipation could in principle
provide very high efficiencies, it is, again, a very uncertain process
and as such provides very little predictive power to allow for
meaningful comparisons with observations.

Synchrotron radiation, with possible self-Compton components, has long
been considered the best candidate to explain the prompt emission of
GRBs. While synchrotron radiation can easily explain the non-thermal
nature of the observed spectrum, more thorough scrutiny reveals
several important problems. First, optically thin synchrotron emission
has a very well defined limiting slope $\alpha\ge-1/3$ (where
$F(\nu)\propto\nu^{-\alpha}$) due to the synchrotron radiation
spectrum of a single electron (the so-called line of death of
synchrotron emission). However, at least several cases of GRBs with
$\alpha<-1/3$ have been detected (Crider et al. 1997; Preece et
al. 1998; Ghirlanda et al. 2002, 2003). Second, the typical slope of
the low-frequency part of the GRB spectrum is $\alpha=0$ (Kaneko et
al. 2006), which is not a natural slope of synchrotron radiation from
shock-accelerated electrons. Third, the high frequency spectrum has a
highly variable slope $\beta$, not easily explained by synchrotron
radiation from shock accelerated electrons that should produce a
fairly standard spectrum with $\alpha\sim1$ (as in the X-ray
afterglows). Finally, for the prompt emission to be efficient,
electrons are expected to cool fast (Ghisellini et al. 2000), but the
typical slope of cooling electrons ($\alpha=1/2$) is not observed in
GRB spectra (Kaneko et al. 2006).

These well known difficulties of synchrotron emission have favored the
study of various alternatives, including quasi-thermal Comptonization
(Ghisellini \& Celotti 1999), jitter radiation (Medvedev \& Loeb 1999;
Medvedev 2000; Workman et al. 2007; Morsony et al. 2008), bulk Compton
(Lazzati et al. 2000); and photospheric emission (M{\'e}sz{\'a}ros \&
Rees 2000; M{\'e}sz{\'a}ros et al. 2002; Rees \& M{\'e}sz{\'a}ros
2005; Pe'er et al. 2005, 2006; Giannios 2006; Giannios \& Spruit 2007;
Pe'er et al. 2007; Thompson et al. 2007; Ryde \& Pe'er 2009, Lazzati
et al. 2009). In this paper we expand the analysis of photospheric
emission as a GRB prompt radiation mechanism, studying the conditions
under which the photosphere can produce a high-frequency component
with a non-thermal power-law shape up to tens of MeV. Photospheric
radiation has two great advantages with respect to synchrotron
emission: it does not require a dissipation mechanism and it naturally
provides high efficiency. Photospheric radiation does not require a
dissipation mechanism if the radiation is released before the full
acceleration of the fireball and therefore at a stage when the
fireball still contains a large fraction of internal energy. Numerical
simulations have shown (Lazzati et al. 2009, hereafter LMB09) that
this is indeed the case for long duration GRBs with a massive star
progenitor. The photospheric efficiency of a typical GRB was found to
be in good agreement with the observations (LMB09; Zhang et
al. 2007). The obvious weakness of photospheric radiation is that it
is customarily assumed to be thermal, therefore lacking the prominent
non-thermal tails observed in GRBs.

The possibility of adding non-thermal tails to the photospheric
spectrum has been investigated before. Pe'er et al. (2006) showed that
continuous dissipation and/or internal shocks at moderate optical
depths give rise to a Comptonized spectrum with a flat energy spectrum
$\nu F(\nu)\propto\nu^0$. Giannios (2006, see also Giannios \& Spruit
2007) performed a similar study and reached analogous conclusions,
focusing on magnetic reconnection as the dissipation mechanism. Here
we argue that a dissipation mechanism that manages to be so continuous
as to keep the electron temperature at a constant equilibrium value
under the intense cooling of IC scattering is very unlikely. A more
realistic assumption is that sub-photospheric electron heating takes
place as a series of one or more episodes of impulsive acceleration
and subsequent cooling. We show that in this case a non-thermal
high-frequency tail is produced, characterized by a slope that is
sensitive to a combination of various parameters and which is
therefore expected to be highly variable during the prompt emission of
GRBs.

This paper is organized as follows: in \S~2 we discuss the origin of
the non-thermal tails, in \S~3 we discuss the propagation of the
non-thermal spectrum in the optically thick sub-photospheric plasma, and
in \S~4 we present Monte Carlo calculations of the model. We summarize
and discuss our results in \S~5.

\section{Sub-photospheric dissipation and spectrum formation}

If dissipation takes place in the highly optically thick phase of the
GRB outflow, multiple scattering and absorption-emission processes are
likely to thermalize the radiation and the baryons, producing a
blackbody radiation spectrum at the same temperature of the baryons
(e.g. Giannios \& Spruit 2007). A more interesting case is the
possibility of dissipation in the sub-photospheric zone, i.e. for
optical depths of several. The sub-photospheric zone of a
long-duration GRB is characterized by the domination of photons over
baryons and leptons, a condition that, as we will see, is fundamental
for producing power-law tails with variable slope. In a spherical or
conical fireball expanding without dissipation (for which
$\Gamma=\Gamma_0 R/R_0$), the comoving\footnote{In this and in the
  next section all quantities are implicitly assumed to be in the
  local comoving frame of the fireball.} baryon density below the
saturation radius reads:
\begin{equation}
n_p=\frac{\dot{M}_{\rm{iso}} R_0}{4\pi R^3 m_p c}
\end{equation}
where $\dot{M}_{\rm{iso}}$ is the isotropic-equivalent mass loss rate
of the central engine, $R_0$ is the nozzle radius, i.e., the radius at
which $\Gamma=1$, and $R$ is the radius of the fireball. The photon
density, assuming the spectrum is a pure blackbody at thermal
equilibrium with the baryons, is given by:
\begin{equation}
n_\gamma=20.2\left(
\frac{L_{\rm{iso}} R_0^2}{4\pi R^4 a \, c}\right)^{3/4}
\end{equation}
where $L_{\rm{iso}}$ is the isotropic-equivalent luminosity of the
central engine and $a=7.56\times10^{-15}$~erg~cm$^{-3}$~K$^{-4}$ is
the radiation density constant. The ratio of photon to baryon
densities below the saturation radius is therefore independent of
distance:
\begin{eqnarray}
\frac{n_\gamma}{n_p}&=&9\times10^{11} \,
\Gamma_\infty\,R_0^{1/2}\,L_{\rm{iso}}^{-1/4}
\nonumber \\
&\sim&3\times10^{5} \,\Gamma_{\infty,3} \, R_{0,7}^{1/2} \,
L_{\rm{iso},52}^{-1/4}
\label{eq:ngne}
\end{eqnarray}
where $\Gamma_{\infty,3}$ is the asymptotic Lorentz factor of the
fireball in units of $10^3$, $R_{0,7}$ is the nozzle radius in units
of $10^7$~cm, and $L_{\rm{iso},52}$ is the isotropic-equivalent
luminosity in units of $10^{52}$~erg~s$^{-1}$. Beyond the saturation
radius, both the baryons and photons densities scale as $R^{-2}$
(since the photon temperature scales as $R^{-2/3}$, e.g.,
M{\'e}sz{\'a}ros \& Rees 2000) and the ratio is again constant out to
the photosphere, where baryon and photons decouple. In any reasonable
condition, the photon to baryon number density ratio of a GRB fireball
is therefore in the hundreds of thousand. In the absence of pairs,
this is also the ratio of photons to leptons. If pairs are present,
the ratio can be altered, but the fact that there are more photons
than leptons holds since only a fraction of all photons have enough
energy to create an electron-positron pair (see below for more
details). Numerical simulations of dissipative fireballs, where the
dissipation is provided by the interaction of the outflow with the
progenitor star, confirm this result.  For example, LMB09 found a
ratio of photons over electrons of $\sim10^6$ at the photosphere in
their simulation of a typical GRB jet with a massive stellar
progenitor.

Under such conditions, a non-thermal spectrum can be generated by
inverse Compton scattering of the thermal photons off mildly
relativistic electrons. The mechanism of the formation of the
non-thermal tail is the following. Consider a dissipation mechanism
that produces a population of mildly relativistic electrons. For
simplicity let us consider a population of thermal electrons with a
typical Lorentz factor $\gamma_e$. The mechanism can heat repeatedly
the same electrons but it is not continuous, so that the electrons
that are heated are allowed to cool before being re-heated. An example
of such a mechanism is repeated shocking by weak shocks, such as those
seen in LMB09. If the photons dominate over the electrons, the
electron cooling is very fast and most of the photons never scatter
off hot electrons, since by the time they scatter off an electron it
has already been cooled through IC interactions with other photons. As
a consequence, the peak frequency of the spectrum is not changed by
the dissipation and subsequent IC process. However, a fraction
$\zeta<1$ of the photons scatter off hot electrons and their energy is
increased on average by a factor $4/3\gamma_e^2$. An even smaller
fraction of photons $\zeta^2\ll1$ is scattered twice by hot electrons
and its energy is augmented by $(4/3\gamma_e^2)^2$. In this way, a
non-thermal tail is built, connected to the peak of the original
blackbody spectrum.

To compute the spectral index $\beta$ of the high frequency
tail\footnote{Note that $\beta$ is defined as the slope of the
  $F(\nu)$ spectrum ($F(\nu)\propto\nu^{-\beta}$), differently from
  the Band $\beta_{\rm{Band}}$ that is defined as the slope of the
  photon spectrum $N(\nu)$. Therefore $\beta=-(\beta_{\rm{Band}}+1)$.}
we proceed as follows.  Let $n_\gamma$ be the comoving density of
photons in the primary blackbody spectrum where the population of
relativistic electrons is embedded. The peak of the $F(\nu)$ spectrum
is proportional to $n_\gamma$, since the total energy density in
photons is $\epsilon\sim n_\gamma h\nu_{\rm{pk}}$ and the peak of the
$F(\nu)$ spectrum is
$F_{\rm{pk}}(\nu)\sim\epsilon/h\nu_{\rm{pk}}=n_\gamma$.

The number of photons scattered is proportional to the number of
relativistic electrons times the number of scattering required to cool
an electron. Since the energy extracted per scattering is on
average\footnote{Note that in this analytical derivation we use
  equations that are formally valid in the $\gamma_e\to\infty$ limit,
  even though we also consider trans-relativistic electrons. In the
  following, a Monte Carlo code will be used to compute spectra more
  accurately.}  $\delta E=4/3\gamma_e^2 h\nu$, the number of
scatterings required to cool an electron is
\begin{equation}
n_{\rm{scatt}}=\frac{3(\gamma_e-1)m_ec^2}{4\gamma_e^2 h\nu}
\end{equation}

Following the above reasoning it's easy to show that $F(\nu_1)\propto
n_en_{\rm{scatt}}$ where $n_e$ is the electron density and $\nu_1$ is
the frequency of the typical blackbody photons that have been scattered once
$\nu_1=4/3\gamma_e^2\nu_{\rm{pk}}$. The power-law spectral slope is
therefore given by:
\begin{eqnarray}
  \beta&=&\frac{\rm{Log}(n_\gamma)-\rm{Log}(n_en_{\rm{scatt}})}
  {\rm{Log}(\nu_1/\nu_{\rm{pk}})}=
  \nonumber \\
  &=&
  \frac{{\rm{Log}}
    \left(\frac{4}{3}\frac{\gamma_e^2}{\gamma_e-1}
      \frac{n_\gamma}{n_e}\frac{h\nu_{\rm{pk}}}{m_ec^2}\right)}
  {{\rm{Log}}(4/3\gamma_e^2)}
\label{eq:slope}
\end{eqnarray}

\begin{figure}[!t]
\plotone{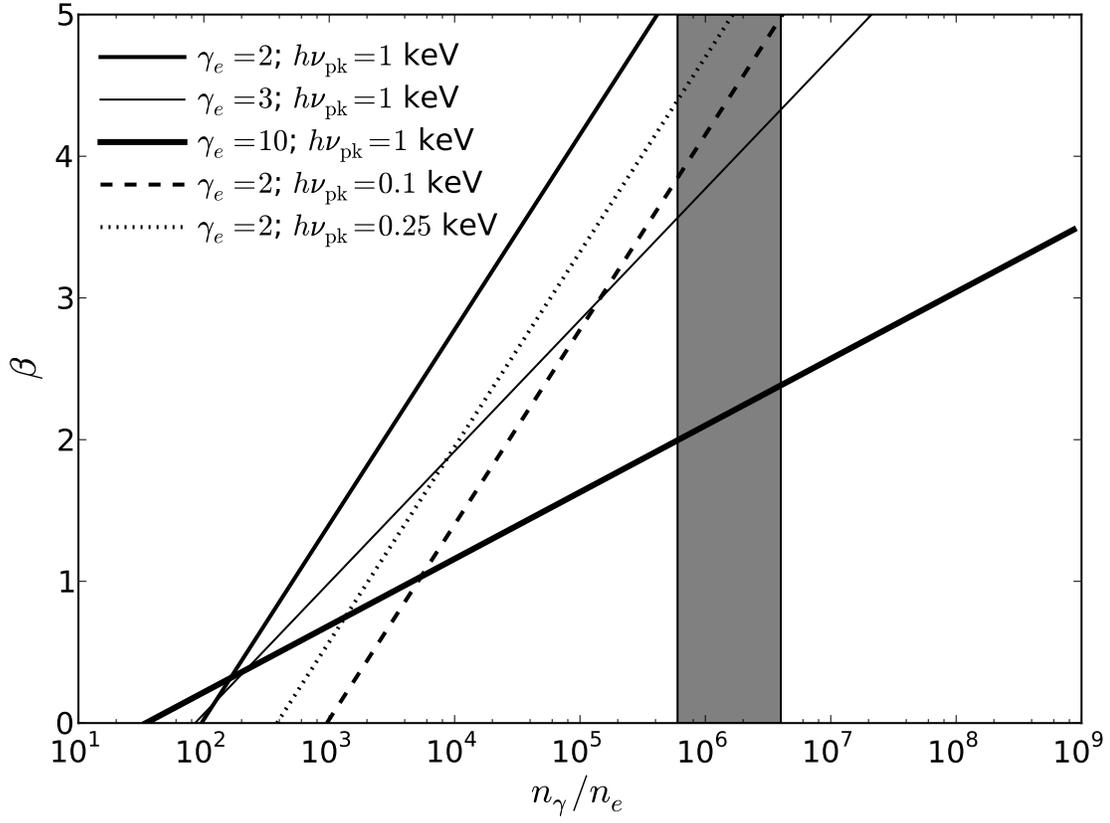}
\caption{{The high energy spectral index $\beta$, defined through
    $F(\nu)\propto\nu^{-\beta}$ plotted versus the ratio of photon to
    electron (or electron-positron pairs) density. Several
    combinations of the parameters $\gamma_e$ and $h\nu_{\rm{pk}}$ are
    shown. The dark band shows the range of ratios observed at the
    photosphere of the LMB09 simulation.}
\label{fig:1}}
\end{figure}

Figure~\ref{fig:1} shows the value of the spectral index $\beta$
as a function of the photon to electron ratio for a sample of
combinations of the remaining two free parameters: the comoving peak
of the blackbody spectrum $h\nu_{\rm{pk}}$ and the typical Lorentz
factor of the accelerated electrons. In all cases the spectral index
grows logarithmically with the ratio of the photon to electron
densities. Therefore, outflows very rich in the radiation component
will have steeper power-law tails. A steepening of $\beta$ is also
observed for increasing comoving peak frequency. Finally, more
energetic electrons give harder tails, at the price of carrying some
bumpiness to the spectrum (see below).

Figure~\ref{fig:1} also shows that there is a limit to the photon to
electron density ratio for which this mechanism is applicable. As we
outlined above, the derivation is based on the assumption that the
probability of a photon scattering off a hot electron is less than
unity. Therefore, the condition of applicability reads:
$n_\gamma/n_e>n_{\rm{scatt}}$ or
\begin{equation}
\epsilon_\gamma>\frac{\epsilon_e}{\gamma_e^2}
\label{eq:condition}
\end{equation}
where $\epsilon_\gamma$ and
$\epsilon_e$ are the energy densities in photons and electrons,
respectively.

\subsection{Repeated dissipation events}

We have so far assumed that there is a single dissipation event that
energizes the electrons that subsequently cool off the blackbody
spectrum, producing a non-thermal high-frequency tail. It is however
possible that the dissipation is intermittent, and the spectrum is
processed by more than one population of hot electrons. For example,
in the simulation of LMB09 at least 6 shocks are detected between the
$\tau_T=4$ region and the photosphere.

If all re-energizations are the same and the electrons always have the
same typical Lorentz factor $\gamma_e$, then the process repeats
identically and repeated accelerations have the same effect as a
lower photon to electron ratio. The spectral slope is therefore
modified as:
\begin{equation}
  \beta_{\rm{multi}}=
  \frac{{\rm{Log}}
    \left(\frac{4}{3}\frac{\gamma_e^2}{\gamma_e-1}
      \frac{n_\gamma}{n_{\rm{acc}}n_e}\frac{h\nu_{\rm{pk}}}{m_ec^2}\right)}
  {{\rm{Log}}(4/3\gamma_e^2)}
\label{eq:slope}
\end{equation}
where $n_{\rm{acc}}$ is the number of acceleration events.

If the different energization events produce electrons with different
typical Lorentz factor $\gamma_e$, and there is one energization that
clearly dominates over the others, then the resulting spectrum is
expected to be dominated by that energization. If, however, the
energizations are similar but not identical to each other, is is hard
to make any quantitative prediction. In \S~4 we show with Monte Carlo
calculations some examples of resulting spectra.

\subsection{Magnetically dominated outflows}

If the outflow is either unmagnetized or moderately magnetized so that
the energy density of the magnetic field is lower than that of the
radiation, the fact that the cooling is dominated by IC scattering is
straightforward. Also in the case of $U_B>U_{\rm{rad}}$, however, the
cooling can be dominated by IC interactions if the peak of the
synchrotron spectrum is self-absorbed (Ghisellini et al. 1998;
Giannios 2008).  Let $\epsilon_B$ be the ratio between the magnetic
energy density and the radiation energy density
$\epsilon_B=U_B/U_{\rm{rad}}$. The intensity of the field is therefore
\begin{equation}
B=\sqrt{8\pi\epsilon_B a}\left(\frac{h\nu_{\rm{pk}}}{2.8 k}\right)^2
\label{eq:B}
\end{equation}
where $k$ is Boltzmann constant. The synchrotron self-absorption
coefficient for a thermal electron distribution is given by
$\alpha_\nu=j_\nu c^2/(2\nu^2 kT)$ (Rybicki \& Lightman 1979), where
the emission coefficient $j_\nu$ at the synchrotron peak frequency is
$j_{\nu_{\rm{syn}}}=0.03 n_e e^3 B/(m_ec^2)$. At the synchrotron peak
frequency $\nu_{\rm{syn}}=0.07 \gamma_e^2eB/(mc)$, the self-absorption
coefficient reads:
\begin{equation}
\alpha_{\nu_{\rm{syn}}}=0.3\frac{n_e}{T\gamma_e^4B}=
5\times10^{-11}\frac{n_e}{\gamma_e^5 B}
\label{eq:alpha}
\end{equation}
where $kT=\gamma_e m_e c^2$ has been used in the right hand
term. Combining Eq.~\ref{eq:alpha} with Eq.~\ref{eq:B}, we obtain:
\begin{equation}
\alpha_{\nu_{\rm{syn}}}=8.7\times10^{-19}\frac{n_e}{\gamma_e^5
  \epsilon_B^{1/2} \nu_{\rm{pk,keV}}^2}
\end{equation}
where $\nu_{\rm{pk,keV}}^2$ is the peak of the thermal spectrum in
units of keV. The optical depth for synchrotron self-absorption at the
synchrotron peak frequency
$\tau_{\rm{syn}}=\alpha_{\nu_{\rm{syn}}}\Delta R$ therefore reads:
\begin{equation}
\tau_{\rm{syn}}=8.7\times10^{-19} \frac{n_e \Delta R}
{\gamma_e^5 \epsilon_B^{1/2} \nu_{\rm{pk,keV}}^2} =
\frac{10^6}{\gamma_e^5 \epsilon_B^{1/2} \nu_{\rm{pk,keV}}^2} 
\label{eq:tausyn}
\end{equation}
where $n_e\Delta R=\sigma_T^{-1}$ at the
photosphere. Equation~\ref{eq:tausyn} gives an optical depth much
larger than unity for moderate values of $\gamma_e$ and for
$\epsilon_B$ not much larger than one, showing that our mechanism is
applicable even to magnetically dominated outflows with
$\epsilon_B\sim100$. High frequency power-law tails from synchrotron
radiation are not expected given the absence of non-thermal electrons.

\subsection{Outside the photosphere}

Even though we have derived the spectrum and applicability conditions
for sub-photospheric dissipation, the unique requirement for this
mechanism to work is that the electrons are cooled by IC scattering
with only a fraction of the photons. This can naturally take place in
the optically thin part of the spectrum. Outside the photosphere the
number of IC scatterings per photon is $<1$, however the number of IC
scatterings per electron is still large. the condition for applicability
of this derivation outside the photosphere reads:
\begin{equation}
\tau_T>\sqrt{\frac{3}{4}\frac{\gamma_e-1}{\gamma_e^2}\frac{m_ec^2}{h\nu_{\rm{pk}}}}
\frac{n_e}{n_\gamma}
\end{equation}
which, according to Eq.~\ref{eq:ngne}, can be a substantially small
number. Therefore, the non thermal spectrum can be generated with this
mechanism out to large radii, approximately 100 times bigger than the
photosphere. This has important consequences for the extent of the
spectrum at high frequencies. If the episodic dissipation that we
envisage here can take place at large radii, well outside the
photosphere, the radiation up to GeV frequencies could be accounted
for.

\section{Radiation transfer in the cooled electrons}

The high frequency power-law tail is produced in a region of
$\tau_T\sim$few. In a relativistically expanding medium, all photons
undergo a number $\sim\tau_T$ of scattering off cold electrons before
leaving the region (e.g., Pe'er et al. 2005). Since the change of
energy of a photon scattering off a cold electron is on average
$\delta \nu/\nu\sim h\nu/m_ec^2$, a cutoff in the spectrum at a
comoving photon energy $h\nu=511/\tau_T$~keV is created.  As long as
the Thomson depth is not too high, the power-law tail is preserved up
to several tens of keV, likely extending into the tens of MeV in the
observer frame.

Another effect of photon propagation, once the non-thermal tail
has been added to the blackbody spectrum, is the creation of
electron-positron pairs via photon-photon collisions. If the pairs
created in such a way do not outnumber the original electrons, the
only effect of their creation is the absorption of all the photons
above $h\nu=511$~keV. This would introduce a cutoff at $h\nu=511$~keV
which is irrelevant since the downscattering off cold electrons
discussed above introduces an even more severe cutoff in the spectrum.
If, instead, the electron-positron pairs outnumber the original
electrons, a pair photosphere is created at a larger radius (Pe'er \&
Waxman 2004; Rees \& M{\'e}sz{\'a}ros 2005). Typically, the pair photosphere
is located a factor 3-10 farther than the electron photosphere. The
processes described in section 2 do not change if the opacity is
provided by pairs rather than by electrons. The number of pairs,
however, is not independent from the number of photons, since the
pairs are created by photon collisions.  The photon to pair
density ratio can therefore be easily calculated:
\begin{equation}
\frac{n_\gamma}{n_{e^+e^-}}=\left(\frac{h\nu_{\rm{pk}}}{m_ec^2}\right)^{-\beta^\prime}
\end{equation}
where $\beta^\prime$ is the slope of the spectrum that created the
pairs and needs not be equal to $\beta$, the slope of the spectrum
created once the pair photosphere has been established. The spectral
slope $\beta$ now reads:
\begin{equation}
  \beta_{e^+e^-}=
  \frac{{\rm{Log}}
    \left[\frac{4}{3}\frac{\gamma_e^2}{\gamma_e-1}
      \left(\frac{h\nu_{\rm{pk}}}{m_ec^2}\right)^{1-\beta^\prime}\right]}
  {{\rm{Log}}(4/3\gamma_e^2)}
\label{eq:slope1}
\end{equation}

\begin{figure}[!t]
\plotone{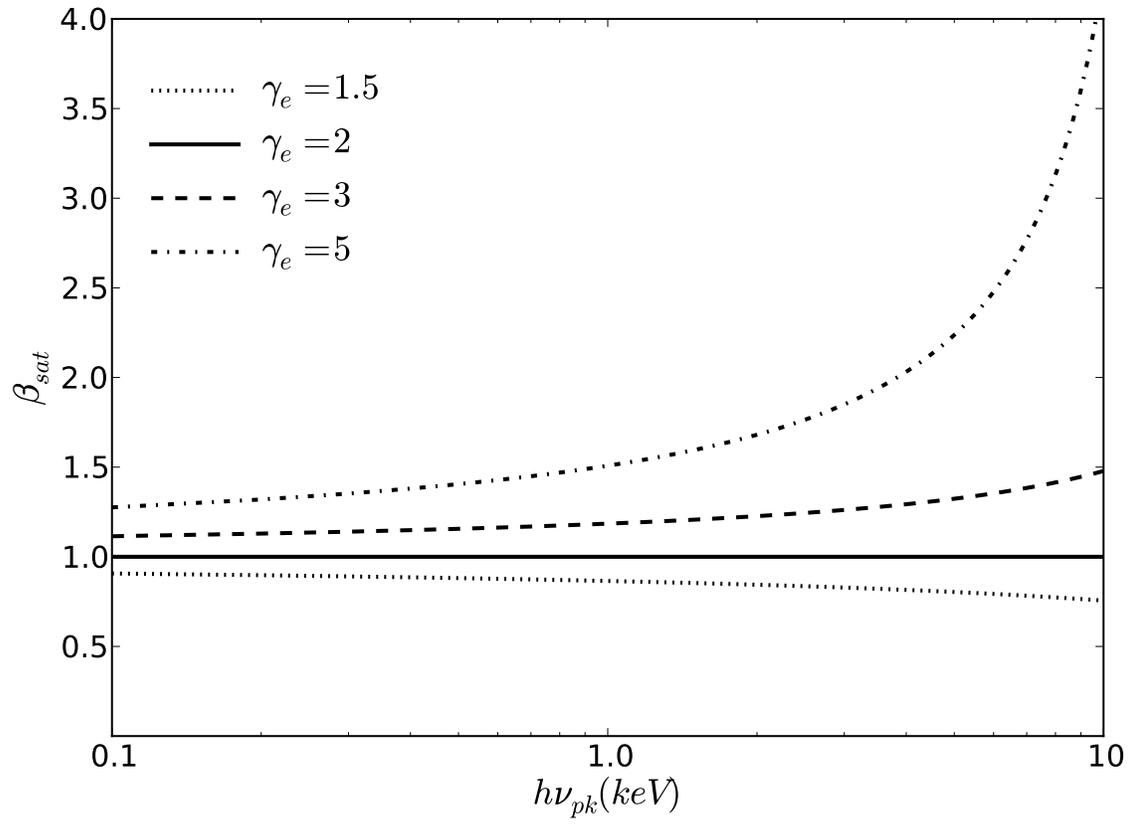}
\caption{{Saturated slope of the high frequency tail of the spectrum
    for a pair photosphere after a few cycles of power-law spectrum
    generation and pair production.}
\label{fig:2}}
\end{figure}

If, however, the cyclical process of creating the power-law tail,
producing pairs from the photons above threshold, producing a new tail
etc. takes place several or more times, it is likely that the spectrum
will saturate such that $\beta_{e^+e^-}\sim\beta^\prime$, resulting in
a saturated slope:
\begin{equation}
\beta_{\rm{sat}}=\frac{{\rm{Log}}\left(\frac{4}{3}\frac{\gamma_e^2}{\gamma_e-1}
\frac{h\nu_{\rm{pk}}}{m_ec^2}\right)}{{\rm{Log}}\left(\frac{4\gamma_e^2}{3}
\frac{h\nu_{\rm{pk}}}{m_ec^2}\right)}
\end{equation}
Figure~\ref{fig:2} shows the value of the spectral slope as a function
of the peak frequency for a range of the electrons' Lorentz
factor. Interestingly, in the vast majority of cases an episodically
dissipative pair photosphere yields a spectral slope
$1<\beta_{\rm{sat}}<1.5$ in fairly good agreement with the
observational results of BATSE (Kaneko et al. 2006).

\section{Monte Carlo simulations}

In the past two sections we have discussed the radiation properties in
the comoving frame of the GRB outflow. In this section we move our
reference frame to the observer and present Monte Carlo (MC)
simulations of the radiation in the observer frame. In this frame, the
spectrum depends on an additional free parameter: the bulk Lorentz
factor of the fireball $\Gamma$. The MC code assumes that a blackbody
spectrum of photons is embedded in a relativistically expanding slab
of electrons with a certain Thomson opacity $\tau_T$. At the beginning
of the simulation, the electrons are accelerated to a unique value of
their random Lorentz factor $\gamma_e$. The position of the photons,
ten million per simulation, is initialized as a random location in the
slab, and after each scattering it is updated according to the
relevant distribution of traveled distances:
\begin{equation}
p(l)\propto e^{-l/l_0}
\end{equation}
where $l$ is the distance traveled by a photon between scatterings and
$l_0=\tau_T^{-1}$ is the mean free path of low frequency photons. The
full Klein-Nishina cross section was used to compute $l_0$ as a
function of the photon energy. At each scattering, the energy of the
photon is modified in two possible ways. If the electron off which the
scattering takes place is relativistic ($(\gamma_e-1)>h\nu/m_ec^2$),
the new photon energy is generated according to the IC energy spectrum
for a single scattering reported, e.g., in Rybicki \& Lightman (1979,
Eq. 7.24). The Lorentz factor of the electron is also updated to take
into account the energy given to the photon. If, on the other hand,
the electron is non-relativistic, Compton scattering is assumed, and
the new photon energy is computed from the Compton equation for a
randomly generated scattering angle (e.g., Rybicki \& Lightman 1979,
Eq. 7.2). Also in this case the energy of the electron is updated
accordingly to the energy loss of the photon.

Adiabatic energy losses are not included in the code because the
cooling time scale of the electrons is much faster than the adiabatic
time scale. The comoving cooling time scale of electrons is 
\begin{equation}
\tau_{\rm{IC}}=\frac{3(\gamma_e-1)m_e c^2}{4\gamma_e^2\sigma_{\rm{T}}c
  aT^4}
\sim 4\times10^{-7} \frac{\gamma_e-1}{\gamma_e^2} \left(\frac{T_{\rm{rad}}}{10^7
    {\rm K}}\right)^{-4} \;\; {\rm s}
\end{equation}
while the comoving adiabatic time scale 
\begin{equation}
\tau_{\rm{adiabatic}}=\frac{R_{\rm{phot}}}{c\Gamma}
\end{equation}
is of the order of seconds. Another effect that is not considered is
the cooling of the photons as a consequence of the expansion of the
electrons (adiabatic cooling of the photons). Due to the small optical
depths considered, this effect is also negligible.

At each step the position of the photons is checked and all the
photons that have reached the surface of the slab are collected in the
output spectrum. The code is stopped when 1/3 of the photons are
collected outside of the slab. The scattering is therefore computed
well beyond the step at which all the electrons have cooled in order
to properly take into account the radiation transfer into the cold
electrons. The output spectrum is then processed to take into account
the fact that the emission is produced on a spherically expanding
surface.

\begin{figure}[!t]
\plotone{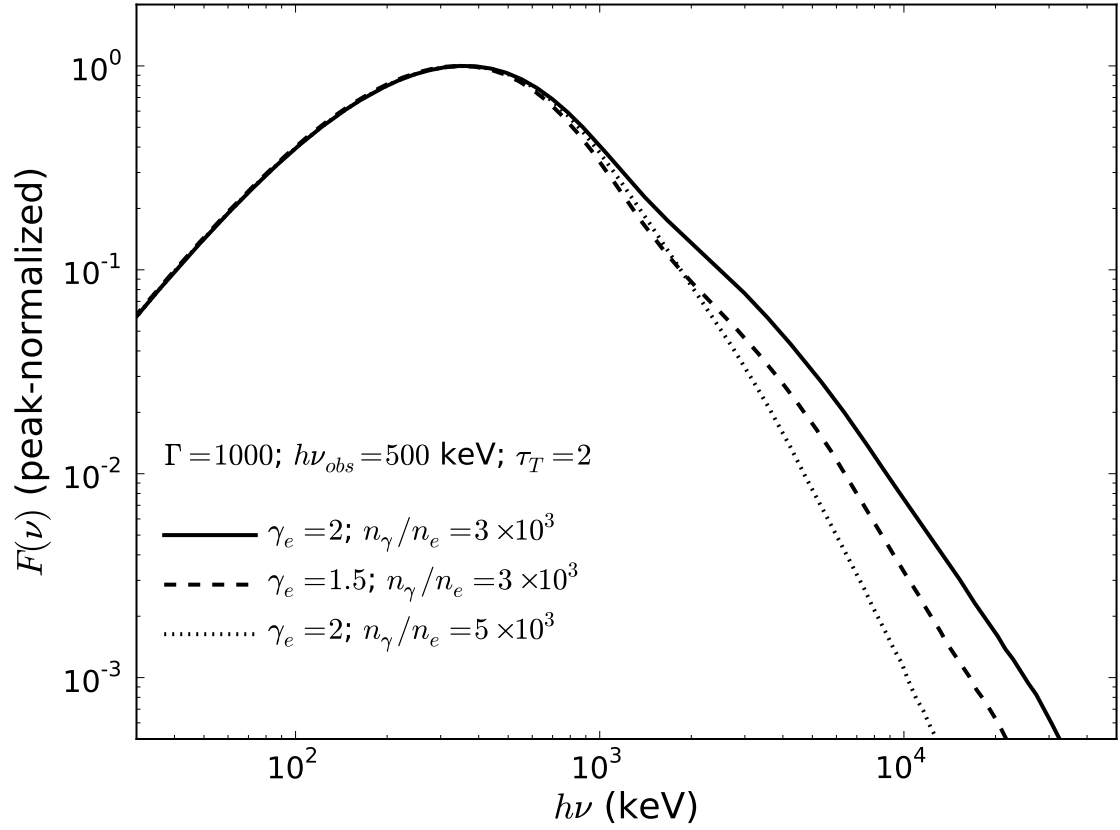}
\caption{{Monte Carlo spectra for a burst with observed peak energy
    $h\nu_{\rm{obs}}=500$~keV and various combinations of the other
    parameters, as indicated in the figure.}
\label{fig:3}}
\end{figure}

\begin{figure}[!t]
\plotone{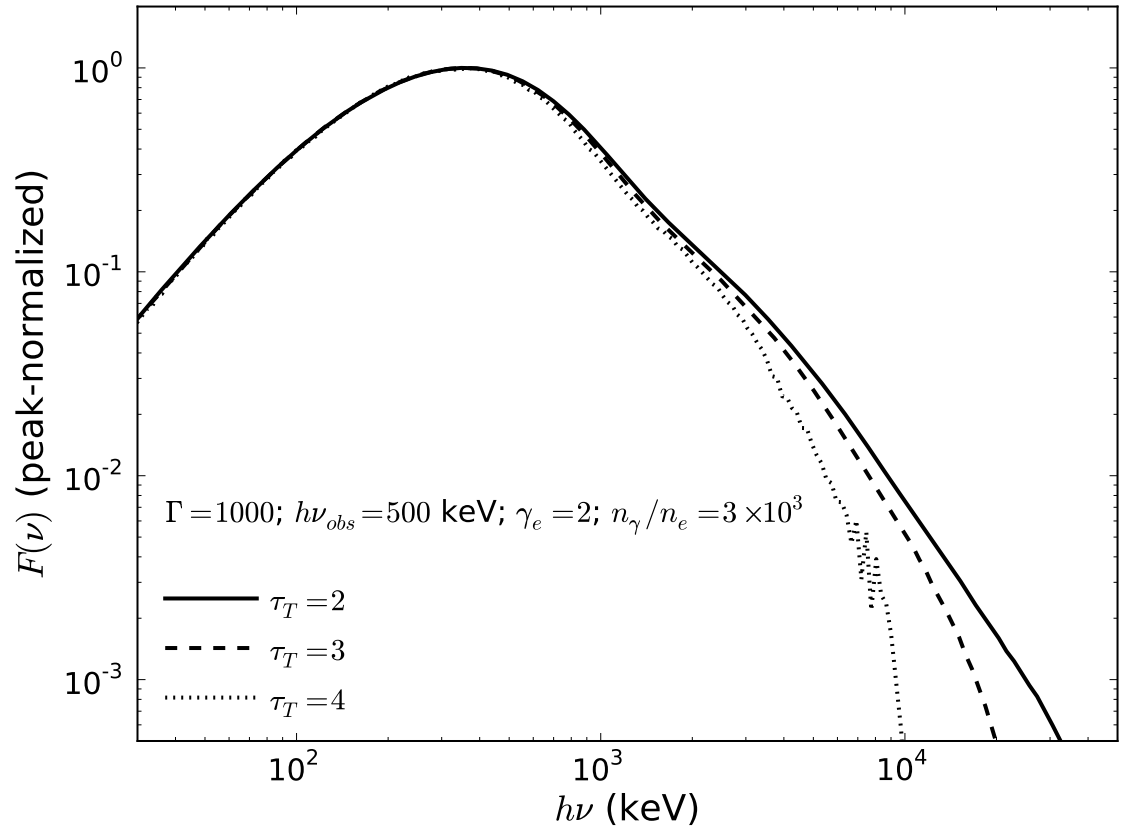}
\caption{{Comparison of Monte Carlo spectra for dissipation taking
    place at different optical depths.}
\label{fig:4}}
\end{figure}

Figure~\ref{fig:3} shows result of three MC runs for a GRB with
observed peak frequency $h\nu_{\rm{obs}}\sim500$~keV and Lorentz
factor $\Gamma=1000$. The three runs are performed for dissipation at
$\tau_T=2$ and with different values of the photon to electron density
ratio and typical electron Lorentz factor $\gamma_e$. The figure shows
that indeed a prominent power-law tail is added to the thermal
spectrum. The tail extends to at least several tens of MeV in the
observer frame, in good agreement with observations. The main
difference with the predictions of \S~2 is that the power-law slope is
steeper by approximately $0.5$.  The predicted slopes are
$\beta=1.23$ (solid line), $\beta=1.53$ (dashed line), and
$\beta=1.98$ (dotted line). However, the slopes measured in the figure
are $\beta=1.72$, $\beta=2$, and $\beta=2.5$, respectively. The reason
for the discrepancy is that the equations of \S~2 assume that the
electrons remain at $\gamma=\gamma_e$ until they cool and
disappear. In reality, photons that scatter more than once off a hot
electron are likely to find the electron slightly cooled at the second
scattering. This has the effect of producing a slightly steeper
power-law slope. As we will see in the following, this can be cured by
assuming that there are more soft photons (so that not all scattering
extract the same energy from the hot electrons). Figure~\ref{fig:4}
shows instead a comparison between the solid line of Fig.~\ref{fig:3}
and two analogous simulations in which the dissipation is assumed to
take place at a higher optical depth $\tau_T=8$ and $\tau_T=16$. As
expected, the three spectra are virtually identical at low frequencies,
but the $\tau_T=8$ spectrum has a cutoff at $\sim40$~MeV and the
$\tau_T=16$  spectrum has a cutoff at $\sim10$~MeV due to
direct Compton scattering.

\begin{figure}[!t]
\plotone{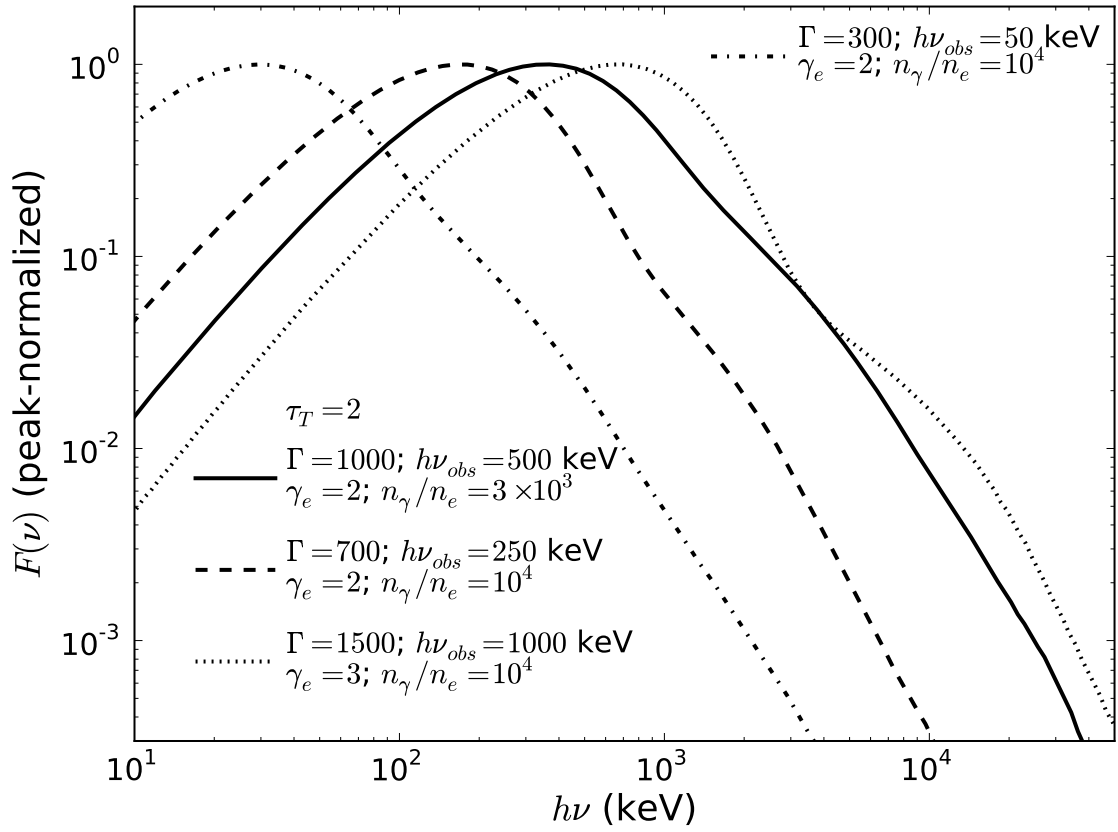}
\caption{{Comparison of Monte Carlo spectra for burst with a wide range
  of peak energy $h\nu_{\rm{obs}}=50$, 250, 500, and 1000 keV. The
  remaining model parameters are indicated in the figure legend.}
\label{fig:5}}
\end{figure}

Figure~\ref{fig:5} shows a comparison of spectra with analogous
characteristics but with different observed peak frequencies and bulk
Lorentz factors. The figure confirms that the mechanism can reproduce
spectra with both low and high peak frequency. An interesting feature
that can be seen in Fig.~\ref{fig:5} is the bumpiness of the dotted
spectrum. It turns out the spectrum looks bumpy for $\gamma_e\ge3$,
i.e., when the photon energy gain per scattering is larger than an
order of magnitude.

\subsection{Repeated dissipation events}

\begin{figure}[!t]
\plotone{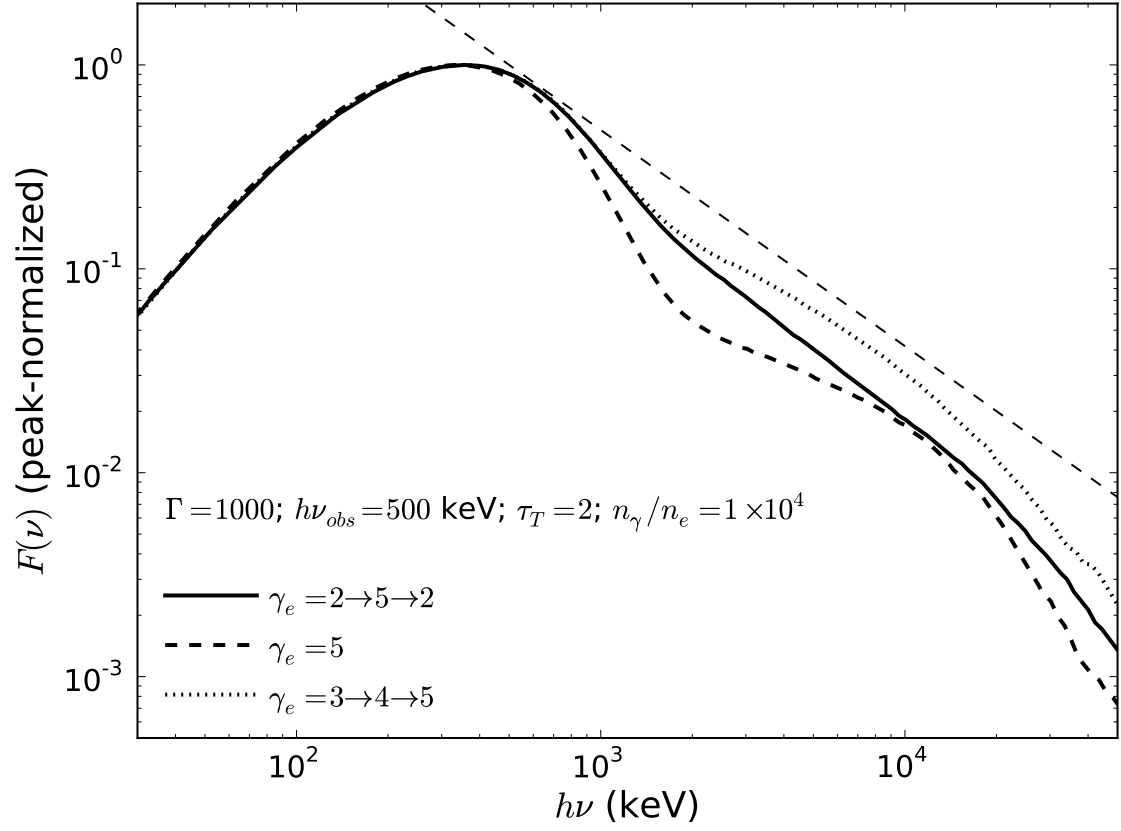}
\caption{{Spectra from multiple dissipation events (the solid and
    dotted lines) compared to the spectrum from a single dissipation
    event (dashed line). The thin dashed line shows the predicted
    slope for the strongest dissipation ($\gamma=5$; $\beta=1.06$).}
\label{fig:6}}
\end{figure}

If the spectrum is produced by multiple dissipation events with
different characteristics (different value of $\gamma_e$), the
analytic predictions do not work and only a numerical calculation can
give the appearance of the final spectrum. Figure~\ref{fig:6} shows the
outcome of two possible situations, one in which there is a
dissipation event which dominates and another in which the three
dissipation events are comparable in strength (and, therefore,
$\gamma_e$). 

The solid line shows the results of a MC simulation in which three
energization events are considered: a small one, yielding
$\gamma_e$=2; an intermediate one, yielding $\gamma_e=5$ and a final
small one, yielding $\gamma_e=2$. As qualitatively predicted, in that
case the spectrum is dominated by the strong event. The main effect of
the smaller events is to pre-broaden the input photon population,
allowing for a smoother power-law rather than a bumpy one (compare to
the dashed line in the same Figure~\ref{fig:6}). The same conclusion
holds even if the three energizing events are characterized by fairly
similar energetics: the resulting spectrum matches fairly well the
prediction for a single energization with the largest $\gamma_e$.

\subsection{Non-thermal input spectrum}

We have so far considered the spectrum arising from the Comptonization
of a primary thermal spectrum. However, observed GRBs not only have
non-thermal high frequency tails, they also have non-thermal
low-frequency tails, with a typical $\alpha=0$ (where, again,
$F(\nu)\propto\nu^{-\alpha}$; Kaneko et al. 2006). Non-thermal
low-frequency power-law tails require a decoupling of the photons from
the electrons below the photosphere and studying the origin is beyond
the scope of this paper. In any case, it is interesting to see what is
the consequence of a non-thermal low-frequency spectrum on the
Comptonized high-frequency tail.

\begin{figure}[!t]
\plotone{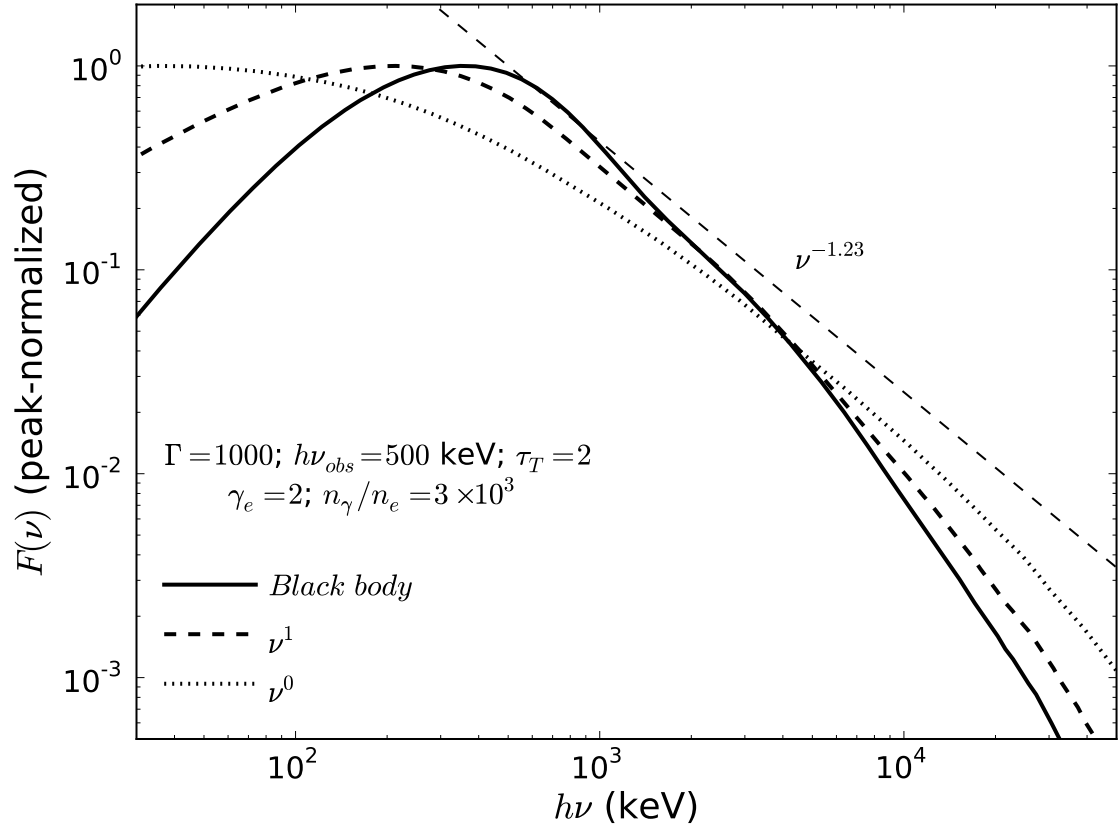}
\caption{{Comparison of Monte Carlo spectra for different primary
    spectra. The solid line is the same spectrum shown in
    Figure~\ref{fig:3}. The figure shows how broadening the
    low-frequency part of the input spectrum results in a better
    agreement with the slope predicted theoretically in \S~2.}
\label{fig:7}}
\end{figure}

\begin{figure}[!t]
\plotone{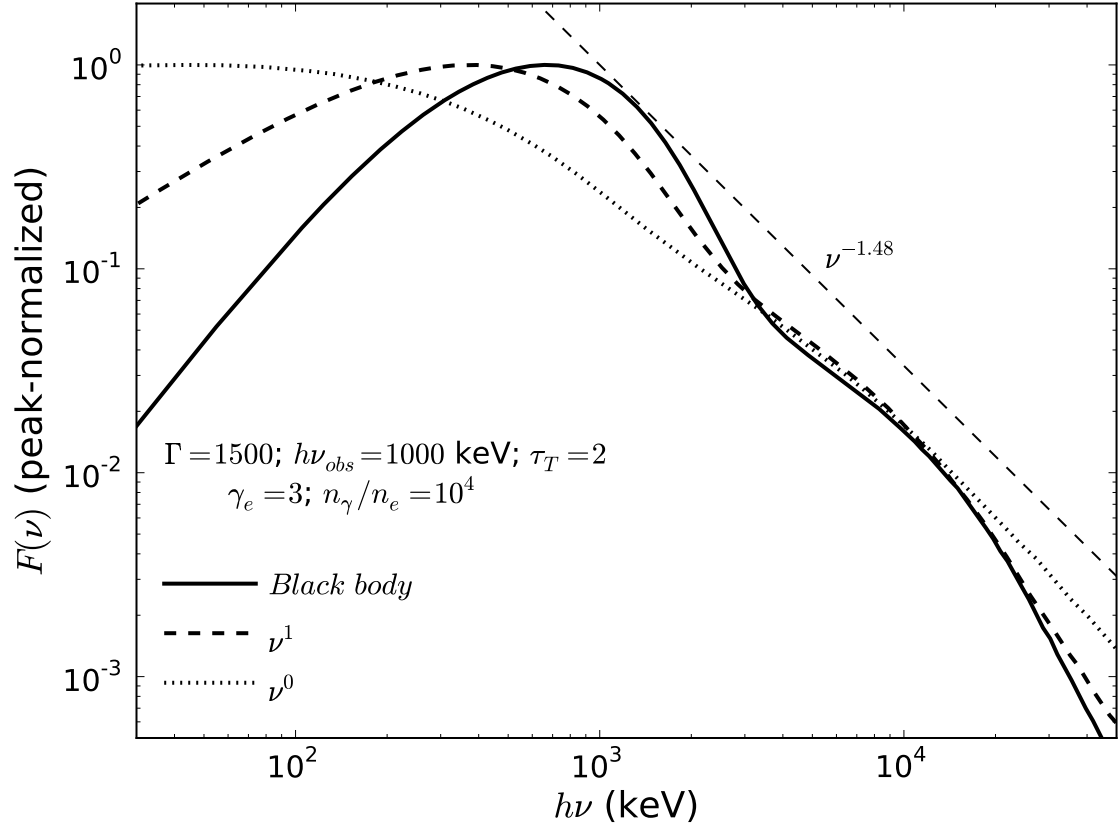}
\caption{{Comparison of Monte Carlo spectra for different primary
    spectra. The solid line is the same spectrum shown with a dotted
    line in Figure~\ref{fig:5}. The figure shows how broadening the
    low-frequency part of the input spectrum results in a smoother
    power-law, without the humps caused by the large energy gain per
    scattering.}
\label{fig:8}}
\end{figure}

To investigate this issue, we performed MC simulations with thermal
spectra modified to have a $\nu^1$ and $\nu^0$ low-frequency
behavior. We find that broadening the input spectrum has two important
consequences. First, the slope of the Comptonized spectrum obtained is
in better agreement with the prediction of Eq.~\ref{eq:slope}. As seen
in Fig.~\ref{fig:7}, the $\nu^1$ input spectrum already hardens the
power-law tail, while the $\nu^0$ input spectrum provides an very good
agreement with the prediction (shown with a thin dashed
line). Figure~\ref{fig:8} shows that besides the better agreement with
the theoretical slope, allowing for a flat low-frequency input
spectrum removes the marked oscillations (or humps) seen in the
high-frequency tails in situations where the energy gain per
scattering is large, again improving the agreement between the spectra
in the figure and the observations.

\subsection{Non-thermal relativistic electrons}

\begin{figure}[!t]
\plotone{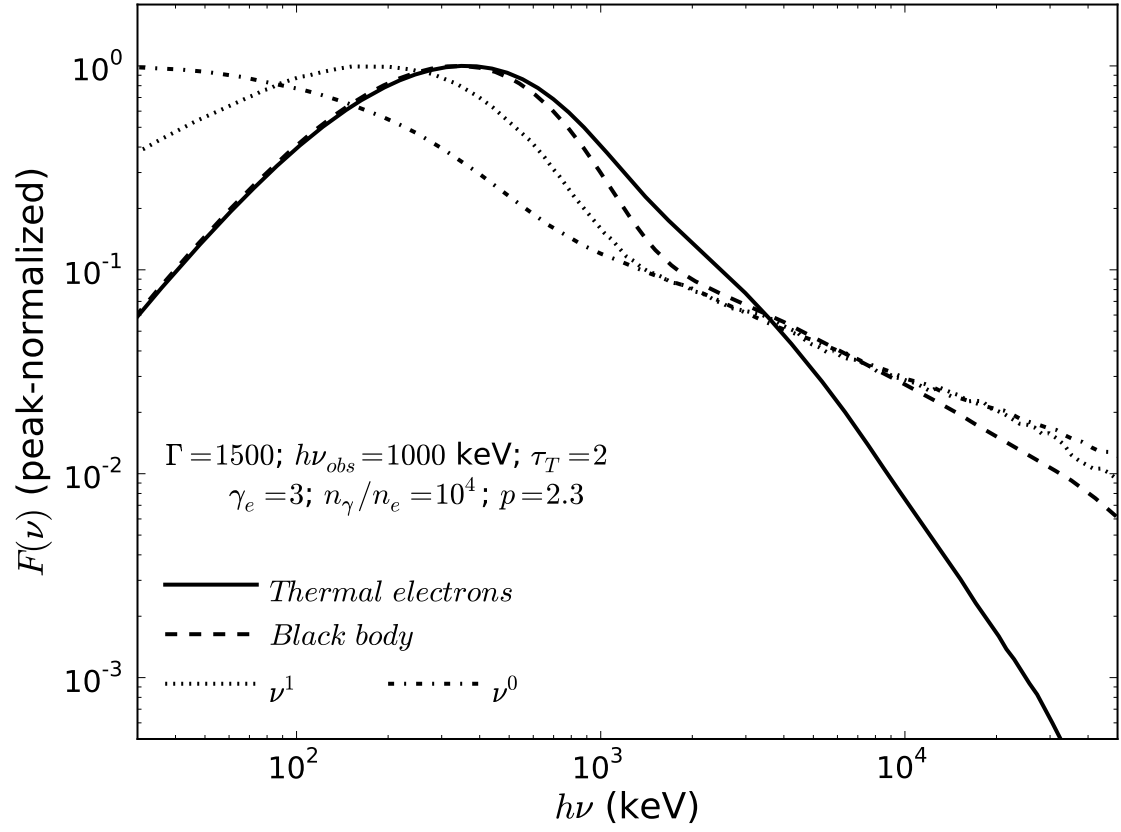}
\caption{{Monte Carlo spectra with non-thermal electron population.}
\label{fig:9}}
\end{figure}

We have so far assumed that the energy dissipation produces a thermal
population of relativistic electrons characterized by a unique Lorentz
factor $\gamma_e$. In some cases, however, dissipation can lead to the
acceleration of non-thermal electrons, with typical energy spectra
$dn_e/d\gamma\propto\gamma^{-p}$, with $p\ge2$. Whether or not the
presence of the non-thermal electrons affect the spectra discussed
above depends on the slope of the electron distribution and on the
fraction of electrons that are contained in the non-thermal tail.  The
strongest influence on the spectrum is observed if all the electrons
are accelerated in the non-thermal tail with a hard typical slope
(e.g., $p\sim2.3$). In that case, the photons that IC scatter off hot
electrons produce a steep non-thermal spectrum with slope
$\beta=(p-1)/2\sim0.65$, $\beta=p/2=1.15$ or $\beta=0.5$, depending on
the cooling regime. Since the majority of the electrons do not scatter
off hot electrons but rather off already cooled electrons, the
power-law tail does not connect smoothly with the primary
spectrum. The results of MC simulations of this scenario are shown in
Figure~\ref{fig:9}. If the primary spectrum is assumed to be a pure
blackbody, the Comptonized one has a prominent power-law tail that
does not connect smoothly to the primary photons, creating a hardening
of the spectrum that is not usually observed in GRBs (e.g. Kaneko et
al. 2006). Only if a very soft input photon spectrum is assumed is a
Band-like spectrum obtained (Fig.~\ref{fig:9}). It is clear, however,
that the mechanism described in this paper produces spectra in better
agreement with observations if the electron acceleration events
contain only a moderate amount of energy and do not produce a sizable
non-thermal tail in the electron population.

\section{Discussion}

We have presented a method by which a primary blackbody spectrum of a
GRB outflow photosphere can be Comptonized into a non-thermal
high-frequency power-law tail. In this scenario, the bulk of the GRB
prompt emission is thermal photospheric and the non-thermal tail
contains a relatively small fraction of the radiation energy. The
majority of the energy of the radiation is therefore released before
it is converted into bulk kinetic energy, solving the problem of
identifying a dissipation mechanism with high efficiency (LMB09).  The
non-thermal tail is produced by thermal electrons accelerated to
mildly relativistic energies by an intermittent dissipation mechanism.
The electrons are allowed to cool through IC interactions with the
photon field between acceleration events. This dissipation mechanism
is not required to have an efficiency larger than a few per cent,
since the power-law tail only contains a fraction of the total energy
of the radiation.The spectral slope of the high energy radiation
depends on the typical Lorentz factor of the accelerated electrons, on
the peak frequency of the thermal photon spectrum, and on the ratio of
the photon to lepton density. Differently from previous work (Pe'er et
al. 2005, 2006; Giannios 2006) we assume that the electrons are
accelerated intermittently and not continuously and that the electrons
are allowed to cool among acceleration events. In addition, we assume
that the energy density in radiation is larger than that in the
relativistic electrons, a situation that is easily realized in jets
born inside massive progenitor stars (LMB09).

Compared to the standard optically thin synchrotron, this mechanism
can explain the steep low-frequency slopes observed in the early
phases of some bright bursts (Crider et al. 1997; Preece et al. 1998;
Ghirlanda et al. 2002, 2003) and the transient thermal bumps detected
is several events (Ryde 2005; Ryde \& Pe'er 2009; Ryde et
al. 2010). Bursts for which no dissipation takes place should not
display any non-thermal features. A complete discussion of the
low-frequency non-thermal tail is however beyond the scope of this
paper and will be presented elsewhere. Explaining a typical
low-frequency tail $F(\nu)\propto\nu^0$ is indeed a challenge for any
model and not only for the photospheric scenarios presented here and
elsewhere (M{\'e}sz{\'a}ros \& Rees 2000; M{\'e}sz{\'a}ros et
al. 2002; Rees\& M{\'e}sz{\'a}ros 2005; Pe'er et al. 2005, 2006;
Giannios \& Spruit 2007; Pe'er et al. 2007; Thompson et al. 2007; Ryde
\& Pe'er 2009, LMB09).

Radiation from dissipative photospheres has been investigated before
(Pe'er et al. 2005, 2006; Giannios 2006; Giannios \& Spruit 2007;
Beloborodov 2010). Giannios (2006) and Giannios \& Spriut (2007)
concentrate on the case of a continuous energy injection or ``slow
heating'', different from our impulsive acceleration assumption. Pe'er
et al. (2005, 2006) consider instead two possible scenarios: ``slow
heating'' and impulsive heating by internal shocks. They find that the
bulk of the photons are shifted in energy once a steady state electron
population is attained. They calculate that, for a fairly wide
parameter range, the comoving peak frequency is of few tens keV.
Attaining such a steady state population requires however a large
optical depth and as a consequence we do not find the same results. To
understand the reason for this, consider an injection episode at an
optical depth of a few. Due to IC scattering, the hot electrons are
efficiently cooled and their energy given to a small fraction of
photons that are shifted to high frequencies. For small optical depth,
the energized photons reach the photosphere with only negligible
losses and decouple from the electron population. The energy given to
the electrons by the dissipation event is therefore given to a small
number of photons that produce a power-law tail as shown in the
figures. In order to obtain a steady state electron population able to
exchange energy effectively with the radiation field, a large number
of scattering per photon is required, so that the high-frequency
photons can return their energy to the electron population. Large
optical depths are therefore required for electron and photon
populations to settle in a steady state configuration dictated by the
balance of heating and cooling. Such optical depths are much larger
than the ones envisioned in this work. In any case, a fraction of the
dissipated energy is left in the electron population that thermalizes
at a temperature higher than the one before the dissipation. For the
parameters adopted in this paper the effect of the increased electron
temperature is negligible because the increase of the electrons
temperature is very small for the optical depths considered.  Finally,
Beloborodov (2010) investigated the radiation produced by a
well-defined dissipation mechanism: nuclear and Coulomb collisions
within a baryonic outflow with a substantial population of neutrons
(see also Beloborodov 2003; Rossi et al. 2006). He finds that a
typical spectrum with a high-frequency slope $F(\nu)\propto\nu^{-1.4}$
is obtained. Differently to our result, the non-thermal spectrum is
due to the presence of non-thermal leptons.

A characteristic of any photospheric model is that the radiation is
released at a small radius $r_{\rm{ph}}\sim10^{10-13}$~cm (e.g. Rees
\& M{\'e}sz{\'a}ros 2005). The main consequence of such a small radius
is that the compactness of the region is large (Pe'er \& Waxman 2004)
and therefore photon-photon interactions result in the production of
electron-positron pairs. This can have two consequences. If the newly
created pairs do not outnumber the original electrons, the only
consequence is the presence of a sharp cutoff in the spectrum at
$h\nu=511$~keV in the comoving frame ($\sim100$~MeV observed). If, on
the other hand, the pairs outnumber the original electrons, a new
photosphere is created, typically at a radius 3 to 10 times larger
than the original photosphere (Rees \& M{\'e}sz{\'a}ros 2005). The
scenario we developed in this paper is not dependent on whether the
photosphere is due to the original electrons or to pairs, the only
difference being that a pair photosphere has a lower photon to lepton
ratio and produces therefore a steeper high-frequency spectrum. In
addition, since the number of pairs and the number of photons are not
independent, we showed that a pair photosphere with at least several
energization events would produce spectra with a typical slope
$1<\beta_{\rm{sat}}<1.5$, in good agreement with observations (e.g.,
Kaneko et al. 2006).

The Fermi satellite has recently shown that at least some bright GRBs
have spectra extending well into the GeV regime (Abdo et al. 2009abc;
2010; Ghisellini \& Ghirlanda 2010; Granot et al. 2010). For any
reasonable combination of parameters, GeV radiation cannot be produced
at the photosphere, since it would be immediately absorbed by
photon-photon collisions to produce pairs. In this scenario, the GeV
emission can be either internal, i.e. produced by dissipation above
the photosphere (see \S~2.2; Toma et al. 2010) or external, due to the
interaction of the outflow with the interstellar medium (Kumar \&
Barniol Duran 2010; Ghirlanda et al. 2010; Ghisellini et al. 2010).

In terms of predictive power, the most notable prediction of this
model is the presence of substantial variability of the high-frequency
spectrum due to the dependence of the spectrum on parameters that are
expected to be highly variable such as, e.g., the peak frequency of
the thermal spectrum in the comoving frame (LMB09). Such variability
might be however hard to detect since it is expected to have short
time-scales and would therefore be averaged out in a time-integrated
spectrum. The dependence of the spectral slope on the peak frequency
of the spectrum opens also the possibility of correlations between the
peak frequency and the spectral slope at high frequencies. Since,
however, the slope depends on two more parameters (the ratio of
photons to electrons and the typical Lorentz factors of the electrons)
it is not obvious that such correlations should be visible in
experimental data.

\acknowledgements We would like to thank Gabriele Ghisellini,
Dimitrios Giannios, and Bing Zhang for useful discussions and
criticisms to an early version of this manuscript. This work was
supported in part by NASA ATP grant NNG06GI06G, Fermi GI program
NNX10AP55G and Swift GI program NNX06AB69G and NNX08BA92G.

\newpage

\end{document}